\documentclass[runningheads]{llncs}

\usepackage{graphicx}
\graphicspath{{./fig/}}
\usepackage{amssymb}
\usepackage{mathtools}
\usepackage{hyperref}
\usepackage{bm}

\begin{document}

\title{Fitting Segmentation Networks on Varying Image Resolutions using Splatting}

\author{Mikael Brudfors\inst{1} \and Ya\"el Balbastre\inst{2} \and John Ashburner\inst{3} \and Geraint Rees\inst{4} \and Parashkev Nachev\inst{5} \and S\'ebastien Ourselin\inst{1} \and M. Jorge Cardoso\inst{1}}
\authorrunning{M. Brudfors et al.}
\institute{School of Biomedical Engineering \& Imaging Sciences, KCL, London, UK \\ \email{mikael.brudfors@kcl.ac.uk} \and
Athinoula A. Martinos Center for Biomedical Imaging, \\ MGH and HMS, Boston, USA
\\ \and
Wellcome Center for Human Neuroimaging, UCL, London, UK
\\ \and
Institute of Cognitive Neuroscience, UCL, London, UK
\\ \and
Institute of Neurology, UCL, London, UK}

\maketitle

\begin{abstract}
Data used in image segmentation are not always defined on the same grid. This is particularly true for medical images, where the resolution, field-of-view and orientation can differ across channels and subjects. Images and labels are therefore commonly resampled onto the same grid, as a pre-processing step. However, the resampling operation introduces partial volume effects and blurring, thereby changing the effective resolution and reducing the contrast between structures. In this paper we propose a \emph{splat layer}, which automatically handles resolution mismatches in the input data. This layer pushes each image onto a \emph{mean space} where the forward pass is performed. As the splat operator is the adjoint to the resampling operator, the mean-space prediction can be pulled back to the native label space, where the loss function is computed. Thus, the need for explicit resolution adjustment using interpolation is removed. We show on two publicly available datasets, with simulated and real multi-modal magnetic resonance images, that this model improves segmentation results compared to resampling as a pre-processing step.

\keywords{Image segmentation \and Splatting \and Resampling \and Pre-processing \and Image resolution.}
\end{abstract}

\section{Introduction}



Automatic semantic segmentation of medical images is widely done using deep-learning-based segmentation networks. To apply these networks, a pre-processing step that resamples all images into the same space is currently performed, as the images can have different orientation, field-of-view and resolution. Choosing the common space can be done in many ways, \emph{e.g.}, based on the median voxel size of the training population \cite{isensee2021nnu}. This step is required for stacking channel dimensions when working on multi-modal(channel) data, but also if a batch size larger then one is required. This type of pre-processing is performed in the majority of biomedical challenges, \emph{e.g}, BRATS \cite{menze2014multimodal}, Medical Segmentation Decathlon \cite{antonelli2021medical} and the WMH Segmentation Challenge \cite{kuijf2019standardized}.

Pre-processing images by resampling to a common space can be seen as a normalisation step, intended to decrease data variance and facilitate both model fitting and generalisability. However, resampling introduces values not present in the original image through interpolation. Furthermore, it has a smoothing effect that, unless coordinates fall exactly at voxel centres, reduces the observed noise variance: let $y = ax_1 + (1 - a)x_2$ be the interpolation of two values $x_1 \sim \mathcal{N}(m_1,~v)$ and $x_2 \sim \mathcal{N}(m_2,~v)$, with $a \in [0, 1]$ ; then, $\text{Var}[y] = (1 - 2a(1 - a))v \leq v$. In addition, interpolation algorithms do not embed prior knowledge about the objects being interpolated, resulting in overly smooth images that can bias analyses and cause false positives \cite{yushkevich2010bias,thompson2011bias}. These limitations can make it challenging to generalise segmentation networks to a wide array of voxel sizes. Furthermore, the networks have no way to know how confident they should be about a particular voxel value, \emph{i.e.}, whether it has been highly interpolated or preserves the raw value. This could be particularly problematic when working with routine clinical MRIs, where thick-sliced (high in-plane, low through-plane resolution), multi-modal acquisitions are the default.

The simplest method for fusing modalities of different image resolution is perhaps to fit separate networks to each modality and then combine their individual predictions. This can be done by integrating multi-modal features from the top layer of each network \cite{suk2014hierarchical,nie2016fully}, or by fusing their outputs via averaging and majority voting \cite{kamnitsas2017ensembles}. However, such output-fusion strategies learn only modality-specific features and ignore the complex interactions between modalities. To better account for correlations across modalities, several layer-level fusion methods have been proposed. For example, Hi-Net \cite{zhou2020hi}, which learns individual modality features via a modality-specific network and a layer-wise fusion strategy, or HyperDense-Net \cite{dolz2018hyperdense}, which employs dense connections across multi-modal streams. However, none of these methods model the fact that the images are defined on different grids in their native spaces. SynthSeg \cite{billot2021synthseg}, on the other hand, introduced a convolutional neural network (CNN) that learns a mapping between multi-modal scans, defined on different grids, by simulating high-resolution scans from the training data by interpolating the low-resolution scans to 1 mm isotropic. Since the interpolation is simulated from the training data, the network becomes robust to variable image resolutions. The method presented in this paper would avoid interpolation, instead using the proposed splat layers. Finally, CNN-based models exist that take irregularly sampled inputs \cite{szczotka2020learning}, but they are currently not easily extended to multi-modal data.

In this paper, we propose a method for directly fitting segmentation networks to the raw image data. Our method is based on the splatting operation, which pushes images, across subjects and channels, into a mean space. The network produces its predictions in this space, which are then pulled back into the native space of each input subject where the loss function is computed. If multiple modalities are provided, the loss is computed on the image on which the target segmentation was annotated. The splat layer avoids interpolating the input data, allowing the network to instead infer on the raw voxel values. We validate our proposed method on two semantic segmentation tasks, on publicly available multi-modal brain MR images. Our validation shows that extending a UNet with our proposed splat layers gives improved segmentations, compared to fitting to data that have been pre-processed by resampling. Our implementation uses a PyTorch backend, with custom splatting and resampling layers written in C++/CUDA, publicly available at \url{https://github.com/balbasty/nitorch}.

\section{Methods}

The idea of our method is quite simple; when fitting a segmentation network, instead of resampling the input images as a pre-processing step, we instead add two new layers to the network, one at its head and one at its tail. These two layers are based on the splatting operation \cite{westover1989interactive}, which allow the network to learn on the raw voxel data of the images. This avoids interpolation that could introduce partial-volume effects, and for the loss to be computed on the native space data. The idea is  that the network implicitly interpolates the data whilst training. To conceptualise the idea of splatting, we next show a simple 1D toy example. The methodology is then extended to $D$-dimensional input tensors in the subsequent section.

\subsection{1D Toy Example}

Let's assume we have a training population of $M$ sets of native-space input vectors ($D=1$), where each set of input vectors represents $C$ channels and can be of different length (\emph{i.e.}, it is not always possible to stack the $C$ vectors of training data). For training a segmentation network, we want to be able to concatenate all input vectors across $C$ and $M$ onto a common grid (\emph{i.e.}, having equal length). Let us define one of these vectors as $\mathbf{f}_{mc}=[10,~11,~12,~13]^{\text{T}}$, with $N_{mc}=4$ elements and the affine mapping\footnote{For medical images, the affine mapping can be read from the image header. In general, it can be defined from knowledge of orientation, pixel size and field-of-view.} $\mathbf{A}_{mc}=\begin{psmallmatrix}2.5 & 0\\ 0 & 1\end{psmallmatrix}$. This vector's identity grid is given by $\mathbf{i}_{mc}=[0,~1,~2,~3]^{\text{T}}$. To resize the input vector on a common training grid, we define its length ($N_t$) and affine mapping ($A_t$). This could be done in a number of ways, in this paper we use a \emph{mean space}. The mean space is defined later; in this example, for simplicity, we assume $N_t=8$ and $\mathbf{A}_t=\begin{psmallmatrix}1 & 0\\ 0 & 1\end{psmallmatrix}$.

There are two ways of resizing the input vector onto the mean-space grid. The standard method is resampling, in which we define an identity grid in the mean space: $\mathbf{i}_{t}=[0,~1,~2,~3,~4,~5,~6,~7]^{\text{T}}$. We then compose the affine mappings as $\mathbf{A}=\mathbf{A}_{mc}^{-1}\mathbf{A}_{t}=\begin{psmallmatrix}0.4 & 0\\ 0 & 1\end{psmallmatrix}$ and transform the identity grid with $\mathbf{A}$ to get the deformation $\bm{\phi}_{t}=[0,~0.4,~0.8,~1.2,~1.6,~2,~2.4,~2.8]^{\text{T}}$. Using $\bm{\phi}_{t}$ we can then pull values from $\mathbf{f}_{mc}$ onto a grid $\mathbf{f}_{t}$ using some form of interpolation. With linear interpolation we get $\mathbf{f}_{t}=[10,~10.4,~10.8,~11.2,~11.6,~12,~12.4,~12.8]^{\text{T}}$. 
This operation can be conceptualised as multiplying $\mathbf{f}_{mc}$ with a sparse matrix $\mathbf{f}_{t}=\mathbf{\Psi}_t\mathbf{f}_{mc}$, where:
\begin{align*}
\mathbf{\Psi}_t = \begin{bmatrix}
1.0, & 0.0, & 0.0, & 0.0\\
0.6, & 0.4, & 0.0, & 0.0\\
0.2, & 0.8, & 0.0, & 0.0\\
0.0, & 0.8, & 0.2, & 0.0\\
0.0, & 0.4, & 0.6, & 0.0\\
0.0, & 0.0, & 1.0, & 0.0\\
0.0, & 0.0, & 0.6, & 0.4\\
0.0, & 0.0, & 0.2, & 0.8
\end{bmatrix}
\end{align*}

\begin{figure*}[t]
\centering
\includegraphics[width=\textwidth]{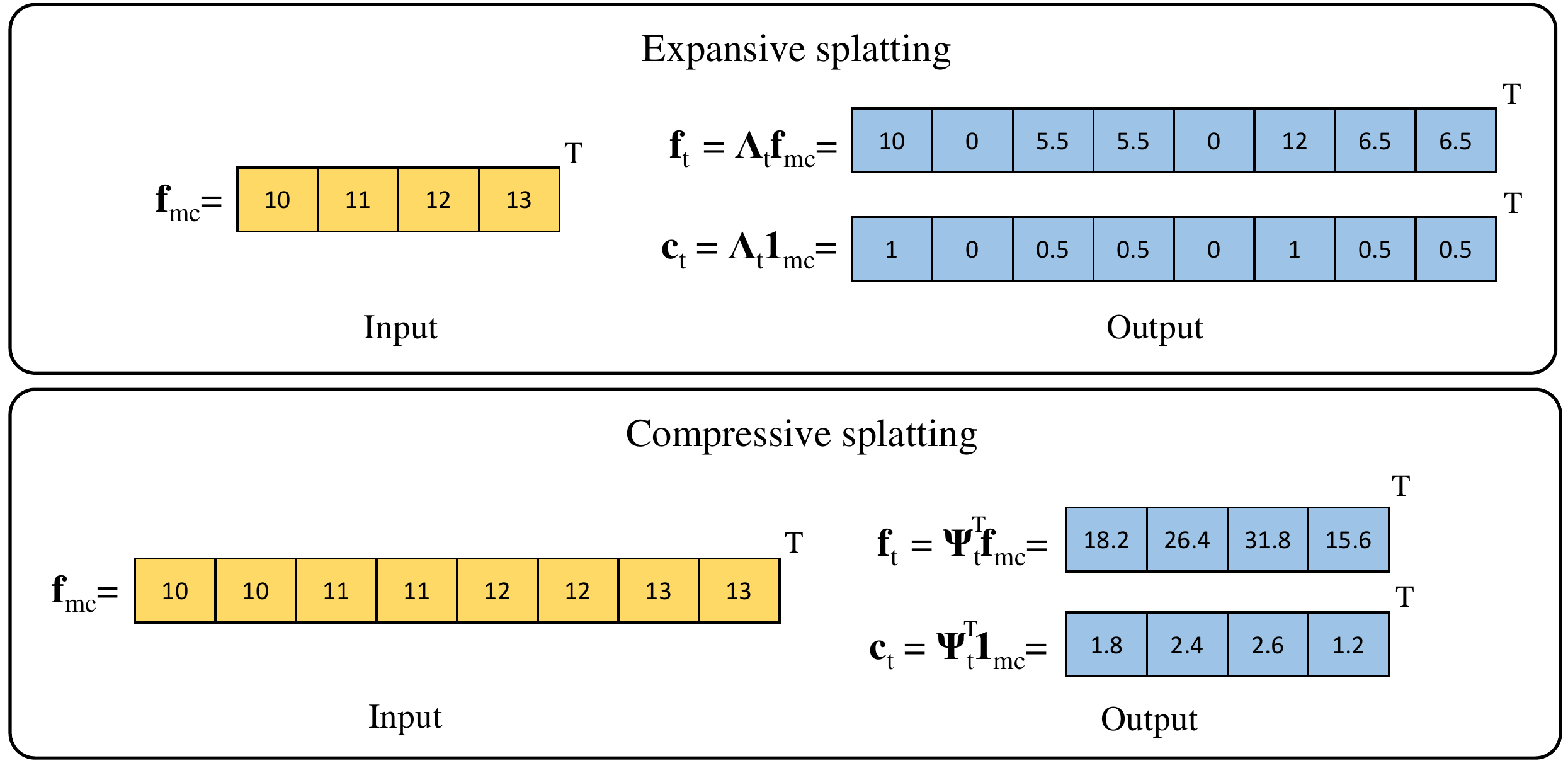}
\caption{1D examples of expansive and compressive splatting.}
\label{fig:splatting}
\end{figure*}

Splatting, on the other hand, does not interpolate the values in $\mathbf{f}_{mc}$. Instead the affine mapping is defined as $\mathbf{A}=\mathbf{A}_{t}^{-1}\mathbf{A}_{mc}=\begin{psmallmatrix}2.5 & 0\\ 0 & 1\end{psmallmatrix}$ and the identity grid of the input vector ($\mathbf{i}_{mc}$) transformed as $\bm{\phi}_{mc}=[0,~2.5,~5,~7.5]^{\text{T}}$. This deformation is then used to push each value in $\mathbf{f}_{mc}$ to a location in the mean-space grid, giving us $\mathbf{f}_{t}=[10,~0,~5.5,~5.5,~0,~12,~6.5,~6.5]^{\text{T}}$. 
As multiple values can be pushed to the same location, a count image is also computed $\mathbf{c}_{t}=[1,~0,~0.5,~0.5,~0,~1,~0.5,~0.5]^{\text{T}}$, which we additionally provide as input to the segmentation network. Note that the splatting is expansive in this example, as there are fewer input voxels than output voxels, so the count image has lots of zeros. Splatting can also be compressive, meaning that multiple values can be pushed onto the same output voxel. As resampling, splatting can be conceptualised as a matrix multiplication $\mathbf{f}_{t}=\mathbf{\Lambda}_t\mathbf{f}_{mc}$, where:
\begin{align*}
\mathbf{\Lambda}_t = \begin{bmatrix}
1.0, & 0.0, & 0.0, & 0.0, & 0.0, & 0.0, & 0.0, & 0.0\\
0.0, & 0.0, & 0.5, & 0.5, & 0.0, & 0.0, & 0.0, & 0.0\\
0.0, & 0.0, & 0.0, & 0.0, & 0.0, & 1.0, & 0.0, & 0.0\\
0.0, & 0.0, & 0.0, & 0.0, & 0.0, & 0.0, & 0.5, & 0.5
\end{bmatrix}
\end{align*}
This matrix is in fact the conjugate transpose of the resampling from the mean space to the input space, that is $\mathbf{\Lambda}_t=\mathbf{\Psi}_{mc}^{\text{T}}$. The expansive and compressive splatting operations are visualised in Fig. \ref{fig:splatting}.

The splatting operation, and its conjugate transpose, are what we add as layers to the head and tail of a segmentation network, respectively. When splatting and its transpose are conceptualised as layers, we now have the first and last layer of the network as adjoint operations, similarly to how convolutions in the encoder and transposed convolutions in the decoder of a UNet are adjoint operations. An additional parallel between splatting (followed by a convolution) and transposed convolutions can be drawn, as they both let the network learn how to optimally invent missing data.

Having the splat operation at the head of the network allows any sized input images to be provided to a network, with multiple channels that may have different sizes or orientations, because the image's voxels will be pushed onto the common training grid. Having the conjugate transpose of the splat operator at the tail of the network allows the loss function (\emph{e.g.}, Dice) to be computed in the image's native space, where the target segmentations were originally drawn. Note that we apply the conjugate transpose to the `logit' output of the network, and afterwards apply the softmax operation. Furthermore, the splatting operations assume that voxels outside the field-of-view are zero, and the conjugate transposes assume that data outside of the field of view are part of the background class. In the next section we extend the operations to an arbitrary number of dimensions.

\subsection{Splatting and Resampling}

Let us write as $\mathcal{N}: \mathbb{R}^{N \times C_i} \rightarrow
\mathbb{R}^{N \times C_o}$ a CNN that maps $N$ voxels and $C_i$ input channels
to $N$ voxels and $C_o$ output channels. Implicitly, each voxel of the grid is
associated with a spatial index $\mathbf{x}_n$, which is identical in the input
and output images, and in all channels within them. Furthermore, voxel
coordinates can be linked to a common coordinate system through the mapping
$\mathbf{A} : \mathbf{x}_n \mapsto \mathbf{y}$, which can be encoded in an
affine orientation matrix $\mathbf{A} \in \mathbb{R}^{D+1, D+1}$ ($D$ is the
dimensionality of the input). These matrices are saved in the headers of 
most medical imaging file formats. In practice, multiple MR contrasts
$\left\{\mathbf{f}_c \in \mathbb{R}^{N_c}\right\}_{c=1}^{C_i}$ may be defined on
different grids, with different coordinate systems $\left\{\mathbf{A}_c \in
\mathbb{R}^{D+1,D+1}\right\}_{c=1}^{C_i}$. Because CNNs require all images to
`live' on the same grid, they are commonly resampled to the same space. For
segmentation, this is typically the space in which the manual labels were drawn.

In general, resampling can be written as $\hat{f}_m = \sum_{n=1}^N f_n 
w(\boldsymbol{\phi}(\hat{\mathbf{x}}_m), \mathbf{x}_n)$, where
$\boldsymbol{\phi}$ is a mapping from the
new grid ($\hat{\mathbf{x}}$) to the old ($\mathbf{x}$) and $w$ is a 
weight that depends on the distance 
between
a voxel $\mathbf{x}_m$ and the sampled location
$\boldsymbol{\phi}(\mathbf{x}_n)$. This 
operation can be conceptualised as a
large linear operation $\hat{\mathbf{f}} = \mathbf{\Phi}\mathbf{f}$, although in
practice, the support of $w$ is small and $\mathbf{\Phi}$ is 
sparse. In this paper, we use trilinear interpolation weights.

Let us write the loss function as $\mathcal{L}$ and the labels as
$\mathbf{f}_l$, the forward pass of the CNN should therefore really be written
as:
\begin{equation}
\left\{\mathbf{f}_c \in \mathbb{R}^{N_c}, \mathbf{A}_c \in \mathbb{R}^{D+1,D+1}\right\}_{c=1}^{C_i}
\mapsto
\mathcal{L}\left(\mathcal{N}\left(\left[\mathbf{\Phi}_c\mathbf{f}_c\right]_{c=1}^{C_i}\right), \mathbf{\Phi}_l\mathbf{f}_l\right)
~,
\end{equation}
although in general the common space is chosen to be that of the labels so that
$\mathbf{\Phi}_l$ is the identity. When labels have a lower resolution than some
of the input images, a different formulation could be to re-slice all images to
the higher resolution space (\emph{e.g.}, 1 mm isotropic), and resample the
output of the network to the label space:
\begin{equation}
\left\{\mathbf{f}_c \in \mathbb{R}^{N_c}, \mathbf{A}_c \in \mathbb{R}^{D+1,D+1}\right\}_{c=1}^{C_i}
\mapsto
\mathcal{L}\left(\mathbf{\Psi}_l\mathcal{N}\left(\left[\mathbf{\Phi}_c\mathbf{f}_c\right]_{c=1}^{C_i}\right), \mathbf{f}_l\right)
~,
\end{equation}
where $\mathbf{\Psi}_l$ maps from the common space to the native label
space, whereas $\mathbf{\Phi}_l$ was used to map from the native label space to
the common space (the underlying transformations $\boldsymbol{\psi}_l$ and
$\boldsymbol{\phi}_l$ are inverse of each other). However, this does not solve
the issues related to the resampling of the input images raised earlier.

In this paper, we propose to replace the initial resampling with the adjoint
operation of its inverse, as part of the forward pass. Since resampling is a linear operation, its adjoint is simply
its transpose $\mathbf{\Psi}^{\text{T}}\mathbf{f}$. In practice, it means that
native data are \emph{splatted} onto the mean space: $\hat{f}_m = \sum_{n=1}^N
f_n w(\boldsymbol{\psi}(\mathbf{x}_n), \hat{\mathbf{x}}_m)$. Importantly,
it means that if the resolution of the common space is higher than that of the
native space, the splatted image has many zeros (the data are \emph{not}
interpolated). The output of the network is then
resampled to the native label space, where the loss is computed:
\begin{equation}
\left\{\mathbf{f}_c \in \mathbb{R}^{N_c}, \mathbf{A}_c \in \mathbb{R}^{D+1,D+1}\right\}_{c=1}^{C_i}
\mapsto
\mathcal{L}\left(\mathbf{\Psi}_l\mathcal{N}\left(\left[\mathbf{\Psi}_c
^{\text{T}}\mathbf{f}_c, 
\mathbf{\Psi}_c^{\text{T}}\mathbf{1}\right]_{c=1}^{C_i}\right), 
\mathbf{f}_l\right)
~,
\end{equation}
where we have let the network know which zeros are missing and which are native 
values, by concatenating splatted images of ones ($\mathbf{\Psi}_c^{\text{T}}\mathbf{1}$) to the input. We note that $\mathbf{\Psi}_c^{\text{T}}\mathbf{f}_c$ can be seen as the gradient of the resampling operation with respect to its input, while $\mathbf{\Psi}_c^{\text{T}}\mathbf{1}$ can be seen as a diagonal approximation of its Hessian.

\subsection{The Mean Space}

What is the best way of defining the common space, in which the training and inference takes place? Using one of the input images to define this space,
for the complete dataset, is not optimal \cite{yushkevich2010bias}. A more principled solution is to compute a mean space from all input orientation matrices
\cite{ashburner2013symmetric,pennec2013exponential}. Briefly, this involves (1)
extracting all linear components from the input orientation matrices; (2)
computing their barycentric mean in an iterative fashion by alternately
projecting them to the tangent space of GL(3) about the current barycentre and
updating the barycentre by zero-centering the tangent data; (3) finding the
closest matrix, in the least square sense, that can be encoded by the product of
a rotation matrix (the orientation) and an anisotropic scaling matrix (the voxel
size). In this work, we compute the mean space once, from the entire training
set, although one mean space per mini-batch could alternatively be used. We
constrain the mean-space dimensions to be a power of two or three, to facilitate
fitting encoding/decoding architectures. Finally, we use a voxel size of 1 mm
isotropic. This could be customised however, \emph{e.g.}, by using larger voxels
for a more lightweight model.


\section{Experiments and Results}

This section investigates whether the splat layer can improve multi-modal MRI brain segmentation in the scenario where, for each subject, we have multiple MR contrasts of differing resolution and the target labels are defined on one of the contrasts. We use a simple baseline network that we fit in two ways: (1) to images that, for each subject, have been resampled to the grid of the target labels; and (2), to native space images, by extending the baseline network with our proposed splat layers. The number of learnable parameters in both networks are the same. 

\subsection{The Baseline Network}

We use a fairly light-weight UNet architecture \cite{ronneberger2015u} with $(16,~32,~64,~128)$ channels in the encoder layer and $(128,~64,~32,~32)$ in the decoder layer, where kernel size $3 \times 3 \times 3$ and stride two is used throughout. This is followed by a layer of $3 \times 3 \times 3$ stacked convolutions with $(32,~16,~16)$ channels each, and a  stride of one. The last layer then outputs the $K$ segmentations labels, which are passed through a softmax. All layers use ReLU activations and batch normalisation. The final network has about 1 million parameters. This is the baseline network, denoted
\emph{UNet}. The UNet is then extended with our proposed splat layers as described in the Methods section. We denote this network \emph{MeanSpaceNet}. The mean-space has dimensions $(192,~192,~192)$ with 1 mm isotropic voxels. Note that the mean-space is defined on only the training data. Both networks are optimised using the Dice loss and the ADAM optimiser (lr=$10^{-3}$). During training, we augment with multiplicative smooth intensity non-uniformities and random diffeomorphic deformations. For the mean-space model, any spatial augmentation needs to be defined in the mean-space and then composed to each image's native space using the affine matrices. We train for a fixed number of 100 epochs, with a batch size of one.

\subsection{Simulated Data: Brain Tumour Segmentation}

\subsubsection{TCGA-GBM Dataset.}

In this experiment, we use the pre-operative, multi-institutional scans of The Cancer Genome Atlas (TCGA) Glioblastoma Multiforme (GBM) collection \cite{bakas2017advancing,bakas2017data}, publicly available in The Cancer Imaging Archive \cite{clark2013cancer}. The dataset was acquired from different MRI scanners. Each subject has skull-stripped and co-registered multi-modal (T1w, T1-Gd, T2w, T2-FLAIR) MRIs and segmentation labels of the enhancing part of the tumor core (ET), the non-enhancing part of the tumor core (NET), and the peritumoral edema (ED). All MRIs have been resampled to 1 mm isotropic voxel size and the same dimensions. In this experiment, we use only the subjects with manually-corrected segmentation labels, which gives in total $N_{\text{gbm}}=97$ subjects, each with four MR modalities and three tumour labels.

\subsubsection{Experiment.}

We simulate two datasets from TCGA-GBM. The first dataset, denoted $\mathcal{D}_{\text{nat}}^{\text{gbm}}$, is created by downsampling the T2-FLAIR image and the segmentation by a factor of three, in a randomly selected dimension. This emulates the situation where manual labels have been drawn on one modality (here the T2-FLAIR), and the other modalities have different voxel size (here the T1w, T1-Gd and T2w). The second dataset, denoted $\mathcal{D}_{\text{res}}^{\text{gbm}}$, is created by trilinearly re-slicing the T1w, T1-Gd and T2w images to the space of the downsampled T2-FLAIR, so that all images have the same dimensions. This in turn emulates the situation where all modalities are resampled to the space of the modality on which the manual labels were drawn. We split the two datasets into equal (train, validation, test) sets as $(40,~17,~40)$. We then fit the UNet to $\mathcal{D}_{\text{res}}^{\text{gbm}}$ and the MeanSpaceNet to $\mathcal{D}_{\text{nat}}^{\text{gbm}}$. Note that it would not be possible to train the UNet model on the $\mathcal{D}_{\text{nat}}^{\text{gbm}}$ dataset, as the subjects' input modalities have different dimensions. After training we apply the two trained networks to their  test sets and compute pairwise Dice scores between predicted and target segmentations.

\begin{figure*}[t]
\centering
\includegraphics[width=\textwidth]{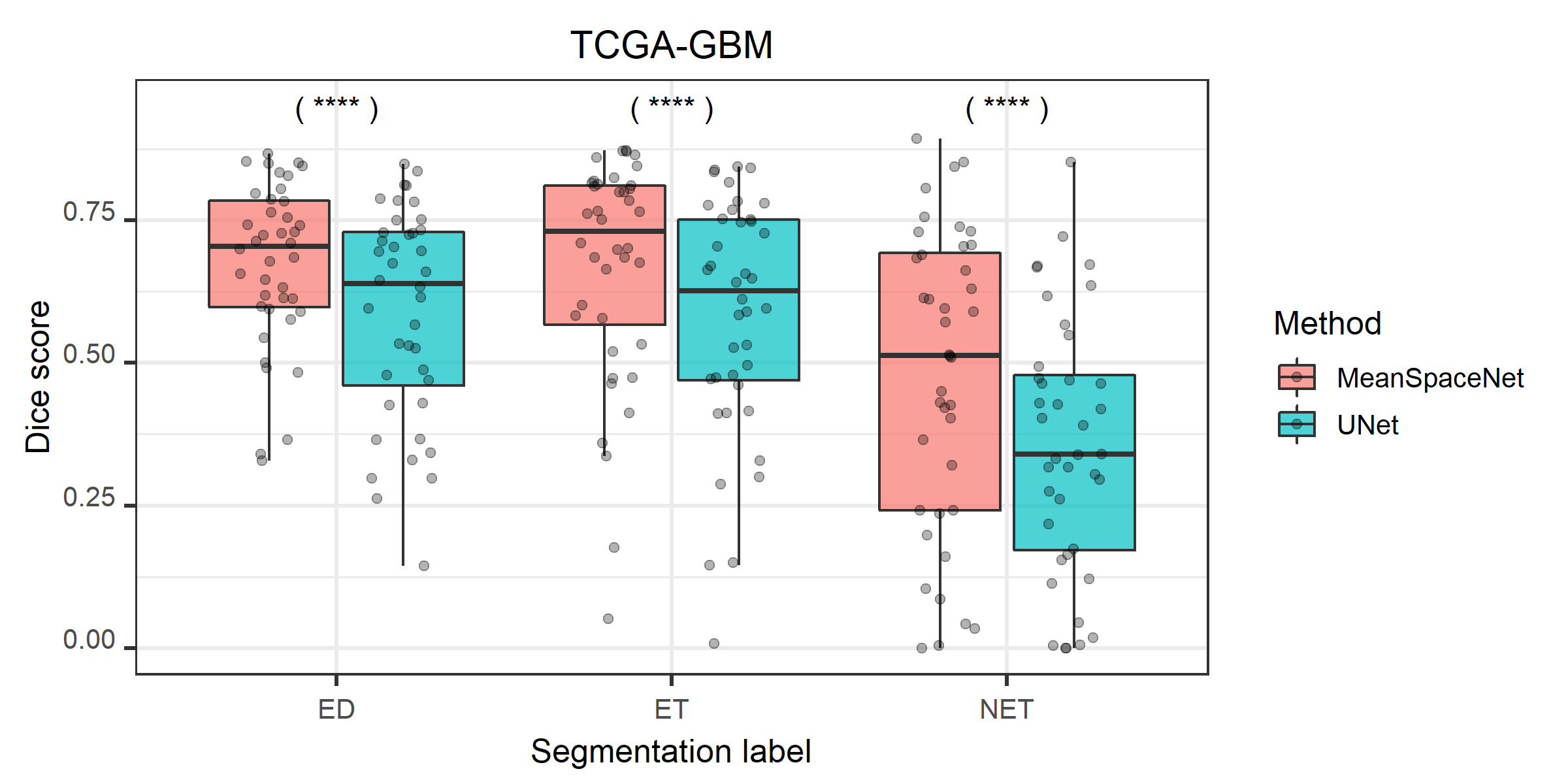}
\caption{Pairwise Dice scores computed on the TCGA-GBM test set ($N_{\text{gbm}}^{\text{test}}=40$), for three tumour labels (ED, ET, NET), and two CNN models (MeanSpaceNet, UNet). On each box, the central mark indicates the median, and the bottom and top edges of the box indicate the 25th and 75th percentiles, respectively. The whiskers extend to the most extreme data points not considered outliers. Asterisks indicate statistical significance of paired Wilcox tests after Holm–Bonferroni correction: 0.05 ($\ast$), 0.01 ($\ast\ast$), 0.001 ($\ast\ast\ast$) \& 0.0001 ($\ast\ast\ast\ast$).}
\label{fig:gbm}
\end{figure*}

\subsubsection{Results.}

The experimental results are shown in the boxplot in Figure \ref{fig:gbm}. The MeanSpaceNet model achieves the best median Dice score over all classes 0.677 vs 0.528; as well as for all individual classes: 0.705 vs 0.639 for ED, 0.731 vs 0.626 for ET and 0.513 vs 0.340 for NET. Paired Wilcoxon tests with Holm-Bonferroni correction shows that the segmentation results are all significant ($p<0.0001$).

\subsection{Real Data: Brain Tissue Segmentation}

\subsubsection{MRBrains18 Dataset.}

In this experiment, we use the original scans (before any pre-processing) from the MICCAI MRBrainS18 grand segmentation challenge (\url{https://mrbrains18.isi.uu.nl}). The dataset was acquired from the same MRI scanner. Each subject has multi-modal (T1w, T1-IR, T2-FLAIR) MRIs and segmentation labels of ten brain structures: Cortical gray matter (CGM), subcortical gray matter (SGM), white matter (WM), white matter lesions (WML), cerebrospinal fluid in the extracerebral space (ECSF), ventricles (VEN), cerebellum (CBM), brain stem (BS), infarction (INF) and other (OTH). In this experiment, we do not use the INF and OTH labels, and we combine WM and WML into a single label. The images' voxel sizes (mm) are: T1w $(1.0,~1.0,~1.0)$, T1-IR $(0.96,~0.96,~3.0)$ and T2-FLAIR $(0.96,~0.96,~3.0)$. The segmentations were drawn in resampled $(0.96,~0.96,~3.0)$ space. In total, we have $N_{\text{brain}}=7$ subjects, each with three MR modalities and seven brain labels.

\begin{figure*}[t]
\centering
\includegraphics[width=\textwidth]{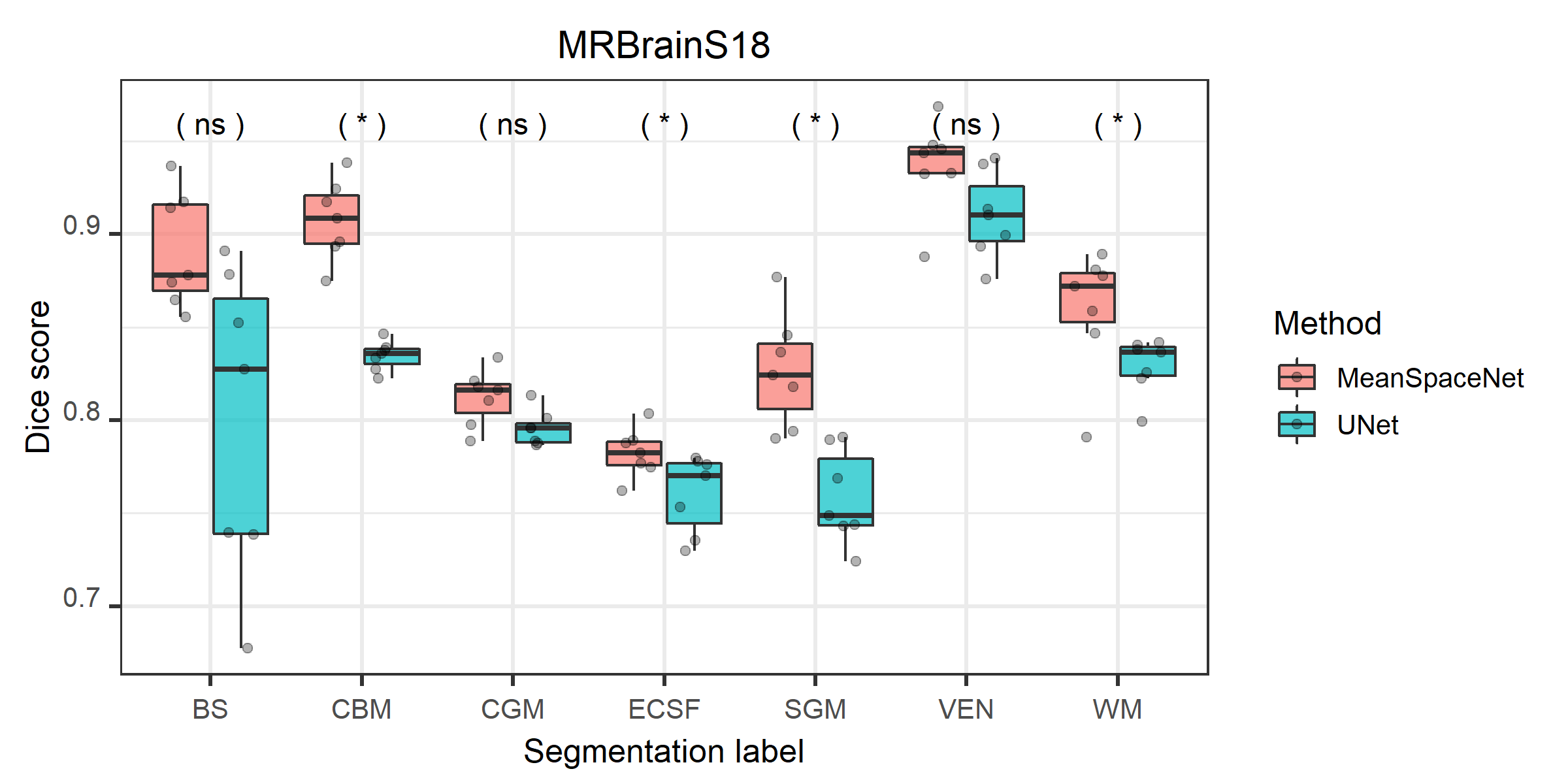}
\caption{Pairwise Dice scores computed on the MRBrainS18 datasets from leave-one-out cross-validation ($N=7$), for seven brain labels (BS, CBM, CGM, ECSF, SGM, VEN, WM), and two CNN models (MeanSpaceNet, UNet).}
\label{fig:brain}
\end{figure*}

\subsubsection{Experiment.}

As in the TCGA-GBM experiment, we construct two datasets. The first dataset, denoted $\mathcal{D}_{\text{nat}}^{\text{brain}}$, is simply the original scans from the MRBrainS18 dataset. The second dataset, denoted $\mathcal{D}_{\text{res}}^{\text{brain}}$, is created by, for each subject in the MRBrainS18 dataset, trilinearly re-slice the T1w image to the space of the segmentation, so that all images have the same dimensions. As the MRBrainS18 dataset has a small number of subjects we perform leave-one-out cross-validation, fitting the UNet on $\mathcal{D}_{\text{res}}^{\text{brain}}$ and the MeanSpaceNet on $\mathcal{D}_{\text{nat}}^{\text{brain}}$, seven times in total, and for each fold computing pairwise Dice scores on the subject held out from training.

\subsubsection{Results.}

The experimental results are shown in Figure \ref{fig:brain}. The MeanSpaceNet model achieves the best median Dice score over all classes 0.864 vs 0.813; as well as for all individual classes: 0.878 vs 0.827 for BS, 0.908 vs 0.836 for CBM, 0.816 vs 0.798 for CGM, 0.783 vs 0.770 for ECSF, 0.824 vs 0.749 for SGM, 0.943 vs 0.910 for VEN and 0.871 vs 0.836 for WM. Paired Wilcoxon tests with Holm-Bonferroni correction shows that the segmentation results are significant ($p<0.05$) for four out of seven segmentation classes.

\section{Conclusions}

In this paper, we described a splat layer that allows a segmentation network to automatically handle resolution mismatches in input data. The idea is to splat each input channel into a mean-space, without interpolating. The forward pass is then computed in this mean-space and its output prediction pulled back into the original resolution of the target labels. The splat layer therefore removes the need for explicit resolution adjustment. We showed on two multi-modal MRI segmentation tasks that splatting was preferred over resampling. Besides allowing segmentation networks to work on the raw image voxels, and computing the loss function in the space of the original target labels, the splat model could also streamline model deployment as a user does not need to ensure that input images have a specific size. The dimension of the mean-space can additionally be defined to allow optimal application of operations, such as strided convolutions, and/or be made small for faster inference. 

Splatting and resampling have the same complexity, which is linear in the number of native voxels. In practice, the loop over native voxels is parallelised (whether running on a CPU or GPU), which makes splatting slightly slower than resampling because multiple native voxels may be pushed to the same output voxel, necessitating the use of atomic assignment (this also makes splatting non-deterministic, as the order in which values are summed-up in an output voxel is architecture-dependent). The cost is therefore somewhat equivalent, compared to resampling all images to the same grid. However, we introduce an additional input channel with the count image, which increases the number of convolution filters in the first layer. However, if the mean space is known a priori, input images can be splatted offline, as a preprocessing step. In this case, only resampling of the prediction to the loss space needs to happen online.



\bibliography{bibliography}%
\bibliographystyle{ieeetr}%

\end{document}